\documentclass[a4paper]{aa}
\usepackage{graphicx}
\usepackage{txfonts}
\usepackage{natbib}

\def\etal{{et al.~}}
\def\ie{{\it i.e.~}}
\newcommand{\kms}{\ensuremath{\mathrm{km\,s}^{-1}}}
\newcommand{\ms}{\ensuremath{\mathrm{m\,s}^{-1}}}

\newcommand{\Uc}{\ensuremath{U_\mathrm{c}}}
\newcommand{\Bc}{\ensuremath{B_\mathrm{c}}}
\newcommand{\Rc}{\ensuremath{R_\mathrm{c}}}
\newcommand{\Ks}{\ensuremath{K_\mathrm{s}}}

\begin{document}

\title{Dust observations of Comet 9P/Tempel 1 \\ at the time of the Deep 
Impact \thanks{Based on observations performed at the ESO La Silla 
     and Paranal Observatories in Chile (program ID 075.C-0583)}}

\author{G.P. Tozzi\inst{1}
          \and
	  H. Boehnhardt\inst{2}
          \and
          L. Kolokolova\inst{3}
          \and
          T. Bonev\inst{4}
          \and
          E. Pompei\inst{5}
          \and
          S. Bagnulo\inst{6} 
 	\and
          N. Ageorges\inst{5}
          \and
          L. Barrera\inst{7}
          \and
	  O. Hainaut\inst{5}
          \and
          H.U. K\"aufl\inst{8}
          \and
          F. Kerber\inst{8} 
          \and
          G. LoCurto\inst{5}
          \and
          O. Marco\inst{5}
          \and
          E. Pantin\inst{9}
          \and
          H. Rauer\inst{10}
          \and
          I. Saviane\inst{5} 
          \and
          C. Sterken\inst{11}
          \and
          M. Weiler\inst{10}
}

\offprints{G.P.~Tozzi ({\tt tozzi@arcetri.astro.it})}

\institute   { INAF - Osservatorio Astrofisico di Arcetri, Largo E. Fermi
             5, I-50125 Firenze, Italy (tozzi@arcetri.astro.it)
	 \and
	     Max-Planck Institute for Solar System Research, 
             Max-Planck-Str. 2, D-37191 Katlenburg-Lindau, Germany,
	 \and
	     University of Maryland, Department of Astronomy, College Park, MD 20742, USA
         \and
             Institute of Astronomy, Bulgarian Academy of Sciences,
             Tsarigradsko chaussee 72, 1784 Sofia, Bulgaria
 	 \and
             European Southern Observatory,
             Alonso de Cordova 3107, Vitacura, Santiago, Chile
         \and
             Armagh Observatory,
             College Hill, Armagh BT61 9DG, Northern Ireland
         \and
             Universidad Metropolitana de Ciencias de la Educacion,
             Av.\ J.P.\ Alessandri 774, Nunoa, Santiago, Chile
        \and
             European Southern Observatory,
             Karl-Schwarzschild-Str.\ 2, D-85748 Garching, Germany
         \and
             Commissariat Energie Atomique
             F-91191 Gif-sur-Yvette, France
         \and
             Deutsche Luft- und Raumfahrt Agentur,
             Rutherfordstr.\ 2, D-12489 Berlin-Adlershof, Germany
         \and
             Vrije Universitet Brussel
             Pleinlaan 2, B-1050 Brussels, Belgium
	  }
\date{Received: today; accepted: tomorrow}


\abstract
{On 4 July 2005 at 05:52\,UT, the impactor of NASA's Deep Impact (DI)
 mission crashed into comet 9P/Tempel 1 with a velocity of about
 10\,\kms.  The material ejected by the impact expanded into the
 normal coma, produced by ordinary cometary activity.}
{Based on visible and near-IR observations, the
  characteristics and the evolution with time of the cloud of solid
  particles released by the impact is studied in order to gain insight
  into the composition of the nucleus of the comet. An analysis
  of solid particles in the coma not related to the impact was also
  performed.}
{The characteristics of the non-impact coma and cloud produced by the
 impact were studied by observations in the visible wavelengths and in the
 near-IR. The scattering characteristics of the "normal" coma of solid
 particles were studied by comparing images in various spectral
 regions, from the UV to the near-IR. For each filter, an image of the
 "normal" coma was then subtracted from images obtained in the period
 after the impact, revealing the contribution of the particles
 released by the impact.}
{For the non-impact coma the Af$\rho$, a proxy of the dust production,
 has been measured in various spectral regions. The presence of
 sublimating grains has been detected. Their lifetime was found to be
 $\sim 11$ hours.  Regarding the cloud produced by the impact, the
 total geometric cross section multiplied by the albedo, SA, was
 measured as a function of the color and time. The projected velocity appeared to obey a
 Gaussian distribution with the average velocity of the order of
 115\,\ms. By comparing the observations taken about 3 hours after the
 impact, we have found a strong decrease in the cross section in $J$
 filter, while that in Ks remained almost constant. This is
 interpreted as the result of sublimation of grains dominated by
 particles of sizes of the order of some microns.}
{}

\authorrunning{G.P.~Tozzi et al.~}
\titlerunning{DI Continuum Imaging at ESO}

\maketitle

\section{Introduction}
The Deep Impact mission (hereafter DI) to the Jupiter family comet
9P/Tempel 1 (hereafter 9P) was aimed at studying the cratering physics in
minor bodies in the solar system and the primordial material preserved
inside cometary nuclei. On July 4, 2005 the impactor of the DI
experiment produced a high-speed (about 10\,\kms) impact in the
nucleus of 9P excavating a considerable amount of cometary
material that was observed and measured both in-situ by the DI fly-by
spacecraft and remotely by Earth-based instrumentation. First results of
the mission are described in \citet{AHearn2005} and
\citet{Sunshine2006}. Early earth-based and other space-based
measurements of the event have been published by \citet{Meech2005},
\citet{Sugita2005}, \citet{Harker2005}, \citet{Lisse2006}, and
\citet{Schleicher2005}.

At the European Southern Observatory (ESO) DI received considerable
observing time allocated to observe the event at their Chilean
observatory, at Cerro La Silla and at Cerro Paranal sites
\citep{Kaeufl2005a}. Here, we summarize results from the visible and
near-IR measurements of the dust in the cometary coma obtained both
shortly before and after the DI event. We focus on the dust ejecta
properties such as scattering properties, projected velocity, and
spatial distribution and their evolution with time.  Complementary
data from the ESO DI campaign on polarimetric and mid-IR observations
as well as on the cometary gas emission and the large-scale coma
activity of the comet are described elsewhere (see e. g., Boehnhardt
et al., 2007). Pre-impact monitoring of the cometary activity is
described by \citet{Kaeufl2005b} and \citet{Lara2006}.

\section{Observations}
\subsection{Telescopes, instruments, filters} 
The majority of the observations, described here, were performed at
the European Southern Observatory (ESO) in La Silla/Chile using the
3.5\,m New Technology Telescope (NTT) by switching between two focal
plane instruments: EMMI (ESO Multi-Mode Instrument), for the visible
spectral region, and SOFI (Son of ISAAC), for the near-IR ({\it
JHK}). Both instruments are of focal reducer type for imaging and
spectroscopic observations.  EMMI provides a field of view of $9.1
\times 9.9$ arcmin with a two-detector array in the red
($400-1\,000$\,nm) and of $6.2 \times 6.2$ arcmin with a single
detector in the blue arm (300-500nm) at 0.32 and 0.37 arcsec/pixel
resolution (using the $2 \times 2$ and $1 \times 1$ binning options),
respectively.  In its large field option used for these observations,
SOFI has a single detector of $4.9 \times 4.9$ arcmin field of view at
0.288 arcsec/pixel resolution. In the visible, narrow band filters,
with bandpasses within selected wavelength regions of interest for
cometary science, were used. In particular, for the study of the
cometary dust, the following filters with no or negligible gas
emission in their passband were used: one in the ultraviolet (\Uc),
one in the blue (\Bc) and one in the red (\Rc) spectral region.  The
near-IR observations were performed with the regular $J$, $H$, and
\Ks\ broad band filters since in this region the gas contamination is
negligible. Table \ref{tablog} gives the log of observations together
with the list of filters used including the respective central
wavelength and full width at half maximum (FWHM) of the wavelength
passband.  Technical information on the La Silla telescope and
instruments can be found at http://www.ls.eso.org/lasilla/sciops.

Since La Silla was clouded over during night 5-6 July 2006, near-IR
imaging of the comet was shifted to the on-going DI campaign at the
Cerro Paranal Observatory using ISAAC (Infrared Spectrometer And Array
Camera) at the 8.2\,m unit telescope Antu of the ESO's Very Large
Telescope (VLT). Due to the shortage of time at the end of the nightly
visibility window, only part of the $J$, $H$ and \Ks\ filter imaging
sequence was performed. ISAAC is a focal reducer-type instrument
providing a field of view of $2.5 \times 2.5$\,arcmin at a pixel
resolution of 0.148 arcsec. Technical information on the VLT and the
ISAAC instrument can be found at http://www.eso.org/paranal/sciops.

\subsection{Calibrations on the sky} 
For calibration purposes, some photometric standards were also
observed before and after the comet observations on each clear
night. In the near-IR, the normal $JH\Ks$ photometric standards were
used \citep{Persson1998}, while for the calibration of the narrow band
filters in the visible, well known spectrophotometric standards from
the list by \citet{Hamuy1994} were measured. The required calibration
frames (bias and sky flatfield exposures for the visible imaging and
screen and/or lamp flatfields with lamp illumination on and off for the near-IR)
were obtained during daytime and/or twilight periods.

\subsection{Observing techniques}
The comet imaging was performed with the telescope tracking at the speed of
the moving target. Jitter offsets of small amplitude
(order of 10-30 arcsec) were applied between individual exposures
through a single filter. As usual for extended objects, the
observations of the comet in the near-IR spectral region were
interlaced by observations of the sky at an offset of $\simeq$ 8$\arcmin$ in a
different region of the sky. A sequence of 5 comet and 5 sky images
were usually taken in each near-IR filter. The jitter sequence typically lasted
for 11-12 minutes per filter. Due to the mentioned shortage
of time, observations on night July 5-6 with ISAAC consisted of only 2
comet and 2 sky images per filter. Observations in the visible were
also repeated 5 times for each filter, offsetting the telescope
by 10-30 arcsec. Calibration observations (standards, sky flatfields)
were performed with the telescope tracking at the sidereal rate. Daytime
calibration images (bias, dome flats) used fixed telescope pointing.

\begin{table*}
\caption{\label{tablog} Filter characteristics and summary log for the comet
observation.  The table lists on the top the filter name, central wavelength,
bandpass FWHM and  the instrument, as well as, in the bottom, the starting exposure
time for the respective filter imaging performed before (- hours) or after (+
hours) the impact. The sky conditions are indicated in the last line of the
table, using the following abbreviations: CLR = clear sky, THN = thin cirrus ,
COUT = clouded out.  On night 5-6/07 the table lists $JH\Ks$ imaging
from the VLT using ISAAC. $t-t_0$ is the time difference between the
observation epoch and the DI impact time at the comet. It is given in hours and minutes,
with negative values before the impact and  positive ones after the event.}
\centering
\begin{tabular}{ccccccccl}
\hline
                         &\multicolumn{6}{c}{BAND}& \\
                         &\Uc    & \Bc  & \Rc   &$J$    &$H$     & \Ks   &  & \\
$\lambda_0$ (nm):        & 372.5 & 442.2& 683.8 &1247   & 1653   & 2162  &  & \\
$\Delta \lambda$ (nm):   & 6.9   & 3.7  & 8.1   & 290   &  297   & 275   &  & \\
Instrument:   		 &EMMI   &EMMI  &EMMI   &SOFI   &SOFI    &SOFI   &  & \\
&$t - t_0$&$t - t_0$&$t - t_0$&$t - t_0$&$t - t_0$&$t - t_0$ & Night         &        \\
&(hh:mm)  & (hh:mm) & (hh:mm) & (hh:mm) & (hh:mm) & (hh:mm)  & (YYYY-MM-DD)  &Weather \\[2mm]
\hline
\hline
&         &         &         &         &         &         &                &       \\
&$-$29:42 &$-$28:13 &$-$28:45 &$-$26:08 &$-$26:21 &$-$26:34 & 2007-07-02/03  &CLR-THN\\
&         &         &         &$-$04:39 &$-$04:52 &         & 2007-07-03/04  &THN \\
&         &         &         &$+$17:29 &$+$17:16 &$+$17:03 & 2007-07/04/05  &CLR \\
&         &         &$+$22:32 &$+$20:29 &$+$17:16 &$+$20:13 &                &CLR \\
&         &         &         &$+$45:32 &$+$45:44 &$+$45:55 & 2007-07-05/06 &COUT \\
&         &         &         &$+$65:39 &$+$65:26 &$+$65:13 & 2007-07-06/07  &THN \\
&$+$90:14 &$+$92:33 &$+$92:21 &$+$94:13 &$+$94:00 &$+$93:47 & 2007-07-07/08  &THN \\
&         &         &         &$+$94:26 &         &         &                &THN \\
&         &         &         &$+$113:55&$+$113:42&$+$113:28& 2007-07-08/09  &CLR \\
&         &$+$118:21&$+$118:40&$+$114:37&$+$114:29&$+$114:09&                &CLR \\
&$+$141:15&$+$138:25&$+$138:12&         &         &         & 2007-07-09/10  &CLR \\
\hline
\end{tabular}

\end{table*}
Since EMMI and SOFI focal plane instruments were mounted on the two
Nasmyth foci of the NTT telescope, fully simultaneous observations in
the near-IR and visible were not possible. However, the switching time
between the two instruments was short (less than 15 min) and allowed us
to use both instruments sequentially during the nightly visibility
window of the comet.

The summary log of observations is
given in Table \ref{tablog}. During the observing period the Sun
(r$_h$) and Earth ($\Delta$) distances of the comet were 1.51\,AU and
0.89--0.91\,AU, respectively. The phase (Sun-Comet-Observer) angle was
$\simeq$ 41\degr, and the position angle of the Sun projected on the
sky at the position of the comet, was $\simeq$ 290\degr.

\section{Data reduction}

\subsection{Frame pre-processing}
For the visible imaging, all comet and standard star images were
corrected for the bias and the flatfield. Both bias and flatfield maps
were computed as the average of a series of bias and sky flatfield
exposures taken during the observing interval and
through the corresponding filters (for the flatfield). Subsequently,
first-order sky background correction was applied by subtracting the
average sky flux value measured at the four edges of the individual
flatfielded images.

For the near-IR data the flatfield maps were computed from the
screen flat images in each filter with the lamp illumination on and
off. Then, for each sequence, a median average sky+bias was computed
from the sequence of five sky observations. The comet images were
then reduced by subtracting the median averaged sky frame and by
dividing the result by the flatfield for the corresponding
filter. Finally, comet images for each filter/sequence in the visible
and near-IR were obtained as the median average of the single 5 images,
after their re-centering on the photometric nucleus. With the median
average of 5 images, all the possible background stars and detector
defects (hot or dead pixels) were almost completely erased. For the
night of July 5-6, this was not possible, since only 2 comet and 2 sky
images were recorded. In this case, the background stars and detector
defects were erased manually. Although for morphology studies this was
acceptable, it prevented any precise quantitative measurements for this particular night.

The same procedure was applied to the standards stars. From the
reduced standard star images, photometric zero points were derived for
clear nights (using aperture photometry and the procedure described in
\citet{Boehnhardt2007} for the EMMI images).

\subsection{Residual background flux removal} 
The presence of a constant residual background was checked and
corrected by measuring the function $\Sigma$Af at large projected
nucleocentric distances, $\rho$. $\Sigma$Af, derived from the Af$\rho$
introduced by \citet{AHearn1984}, describes the dust albedo (A)
multiplied by the total area covered by the solid particles in an
annulus of radius $\rho$ and unitary thickness. It is equal to $2\pi
\rho$Af, where f is the average filling factor of the grains at the
projected distance $\rho$. Note that the definition given here is
slightly different from the the original one given in \citet{Tozzi02}
and \citet{Tozzi04}, even though the physical meaning is the same.
Assuming a simple outflow pattern, i.e. geometric attenuation and
expansion at constant outflow velocity of the cometary dust, the
$\Sigma$Af function should be independent of $\rho$.  In this case, a
small residual background (for instance, from incomplete sky
subtraction) would introduce a linear dependence of the function with
$\rho$. Hence, by applying a trial and error procedure, the residual
background can be removed from the reduced images such that the
$\Sigma$Af function becomes constant at large $\rho$.  This procedure
does not affect the detection of changes in the cometary activity,
since the latter introduces an ``expanding bump" in the profile (see
section 4.2.2), a very different behavior from the linear dependence
with $\rho$~ introduced by uncorrected background subtraction.  This
approach for residual background removal is still applicable for the
observations taken within about a day after the impact, since the dust
produced by the impact was confined to cometary distances shorter than
20\,000\,km and the coma flux measured in the SOFI images at larger
distances could still be used for the above-mentioned
calculations. This method of the residual background removal is not
easily applicable to the coma images taken with ISAAC during the night
July 5-6, 2005 since the DI ejecta had already expanded to the edge of
the field of view. For those observations, we assumed that the
integral over the position angle (PA) of the function $\Sigma$Af,
obtained on the opposite side of the ejecta cloud (over PA between 0
and 90 deg), remained unchanged from night to night. By changing the
background level in ISAAC images, the flux profiles measured in this
quadrant before and after DI were forced to be constant with
$\rho$. Indications for the assumption of unchanged appearance of the
normal coma comes from the analysis of pre-impact observations
\citep[see][]{Lara2006} and from the fact that the coma signal
disappears in the respective PA range when subtracting a pre-impact
image from a post-impact one (both taken through the same filter).

\subsection{Flux calibration} 
All the images taken in clear conditions were then calibrated in
Af using the following formula, derived from \citet{AHearn1984}
\begin{equation}
{\rm Af}=5.34 \times 10^{11} \left(\frac{r_h}{dx}\right)^2 C_s \times 10^{-0.4(Z_p-M_s)}
\end{equation}   

where $r_h$ is the heliocentric distance in AU, $dx$ is the detector
pixel size in arcsec, $C_s$ is the pixel signal in $e^-$/s, $Zp$ and
$Ms$ are the zero points and the solar magnitude in the
used filter, respectively.  Images taken at non-photometric conditions
are flux calibrated assuming that the $\Sigma$Af profiles at large
$\rho$ are coincident with those of the day before and/or day
after. This was justified by the fact that, due to the low dust
expansion velocity, any change in the dust production would not affect
regions at $\rho$ larger than 20\,000--30\,000\,km in 24
hours. Moreover, the coma analysis by \citet{Boehnhardt2007} suggests
that no significant changes in the flux distribution (except for the
DI ejecta cloud) took place between July 3 and 10, 2005.  The relative
calibration from consecutive good nights was checked by comparing
the $\Sigma$Af values of the comet at large nucleocentric distances.

By a careful examination of all comet exposures, we noticed that the
images recorded before and 90 hours after the impact were very similar
and no or negligible traces of DI ejecta were found.  Hence, in order
to increase the signal-to-noise (SN) ratio, we computed images (one
per filter) of the "undisturbed" comet (hereafter called the "quiet" comet) as
the median average of the comet images taken
before and after 90 hours from the impact. The standard deviation of
the median average is within $2-6$\,\% for the most part of the comet,
\ie, for regions at nucleocentric distances between 2\,000 and
50\,000\,km. In regions closer to the nucleus, this standard deviation
increases slightly because of the effect of different seeing in the
various nights. It also increases at distances larger than 50\,000\,km
due to the low coma signal.

The subsequent scientific analysis of the calibrated images is based
mainly on $\Sigma$Af profiles and Af$\rho$, both easily obtained by
numerical integration of the flux in the comet images in concentric
apertures centered on the nucleus. The physical meaning of the
$\Sigma$Af profiles is described above. Following the original
definition by \citet{AHearn1984}, Af$\rho$ is proportional to the
average comet flux in the aperture multiplied by its equivalent
cometocentric projected distance $\rho$. This function does not depend
on $\rho$ for constant outflow velocity. Thus, when using filter
images taken in the dust continuum bands, Af$\rho$ is a proxy of the
dust production rate, $Q_\mathrm{dust}$, of the cometary
nucleus. However, due to unknown dust properties such as the dust size
distribution and the dust albedo A, it is not straightforward to
quantify $Q_\mathrm{dust}$ using Af$\rho$ measurements of comets.

\section{Results}

\subsection{The "quiet" comet}
\subsubsection{Af$\rho$ and $\Sigma$Af as a measure of cometary dust production}
\begin{figure*}
  \begin{center}
\resizebox{!}{4.5cm}{\includegraphics{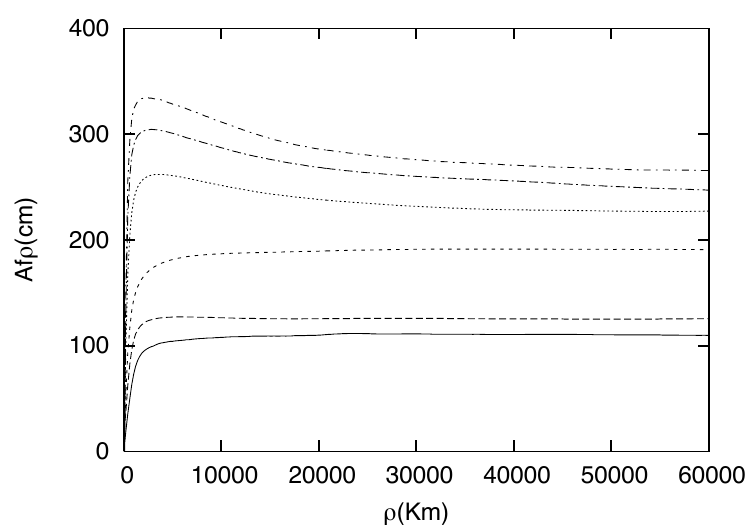}}
\resizebox{!}{4.5cm}{\includegraphics{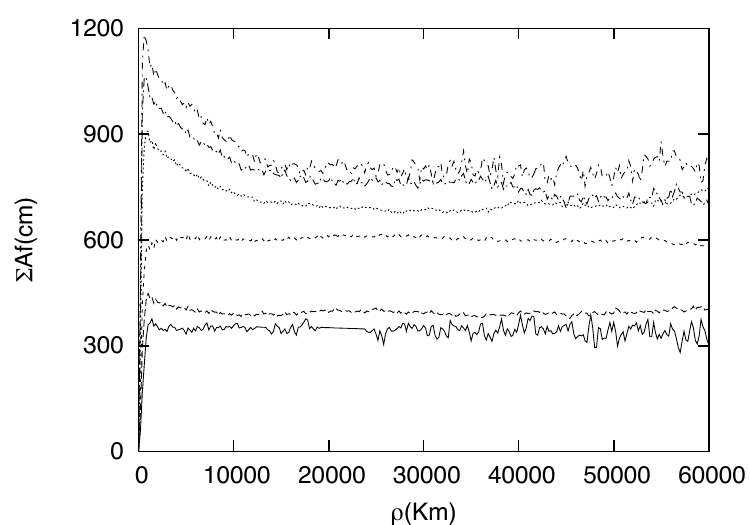}}
\caption{\label{figqc2} Af$\rho$ and $\Sigma$Af profiles determined from the "quiet"
comet (see text): \Uc\ (solid line), \Bc\ (long dashed line), \Rc\
(short dashed line), $J$ (dotted line), $H$ (Long dashed - dotted
line) and \Ks\ (short dashed - dotted line) }
\end{center}
\end{figure*} 
The $\Sigma$Af and Af$\rho$ profiles of the "quiet" comet, derived
from the observations of 9P in the six continuum bands \Uc, \Bc, \Rc,
$J$, $H$, \Ks\ during the nights on July 2/3, 3/4, 7/8, 8/9 and 9/10,
2005, are plotted in Fig. \ref{figqc2}. Various pieces of information on the
comet dust production can be derived from these profiles. The
horizontal profiles at distances beyond 10\,000\,km from the nucleus
suggest a steady-state level in the dust production that resembles
homogeneous and isotropic dust expansion in the coma at a constant
speed. From the existence of jet and fan structures in the 9P coma and
since the radiation pressure modifies the dust distribution in the
coma, it is clear that these ideal conditions are not
fulfilled. However, as long as the jets and fans are stable, \ie they
don't change the dust production, the Af$\rho$ and $\Sigma$Af
functions are constant. The solar radiation pressure may introduce a
linear dependence of these functions with $\rho$, but it becomes
noticeable only at large scales.  The Af$\rho$ values as a measure of
the dust production of the "quiet" comet are determined at projected
distances larger than 40\,000\,km. where the radial profiles in
Fig. \ref{figqc2} have reached constant values. Results are presented
in Table \ref{tabAfrho}.

\begin{table}
\caption{\label{tabAfrho} Measured Af$\rho$ for the "quiet" comet (see text)}
\centering
\begin{tabular}{cc}
\hline
Filter & Af$\rho$(cm) \\ 
\hline
\Uc & $111 \pm 11$ \\ 
\Bc & $125 \pm 12$ \\ 
\Rc & $191 \pm 19$ \\ 
$J$ & $228 \pm 23$ \\ 
$H$ & $253 \pm 25$ \\ 
\Ks & $269 \pm 27$ \\ 
\hline
\end{tabular}
\end{table}

The error in the Af$\rho$ measurements is mainly due to the relative
photometric calibration error, which is estimated to be of the order
of 10\,\%. The Af$\rho$ values given here are slightly higher than the
value of 112 cm given by \citet{Schleicher2005}~ for observations in
the green wavelengths (445--526\,nm). They are also higher than the
value of 102 cm, later revised to 99 cm, derived from Rosetta/Osiris
observations, using the NAC (Near Angle Camera) broad-band filters
(\citep{Keller2005} and \citep{Keller2007}).  However, as already
pointed out by Schleicher et al., this may be due to the larger phase
angle of the spacecraft observations (69\degr) compared to the
measurements from the Earth (41\degr).

\subsubsection{Signatures of dust sublimation:}
It is evident from Fig.~\ref{figqc2} that the near-IR $\Sigma$Af
profiles are not constant. They increase significantly for distances
smaller than $\approx$ 15\,000\,km, showing also a little spike very
close to the photometric nucleus. The spike is probably the signature
of the nucleus convolved with the seeing. However, the SN ratio of
this signature is too low to derive any useful information. Instead,
it can be evaluated through the spatial resolution imaging of the coma
using adaptive optics systems, such as those collected
during the impact week using the NACO instrument at the VLT (not
described here). The slow increase of $\Sigma$Af cannot be due to
dynamical phenomena (for example, increased cometary activity), since
the near-IR profiles derived from different observing nights look very
much the same. Instead the $\Sigma$Af profiles in the visible do not
show any evident increase at small nucleocentric distances.

Similar $\Sigma$Af profiles have been found in comet C/2000 WM$_1$
(LINEAR) (hereafter WM$_1$) \citep{Tozzi04}. At that time this
phenomenon was interpreted as a result of sublimation of two kinds of
organic grains: one with a lifetime of $\approx$1.3 h and the other of
$\approx$ 17 h. Following the analysis of the WM$_1$ data, the near-IR
$\Sigma$Af profiles of 9P for $\rho >$ 1\,000\,km were fit by a
function of the kind
\begin{equation}
 \Sigma {\rm Af}(\rho) = \Sigma {\rm Af}_0+ \Sigma {\rm Af}_1 \times e^{-(\frac{\rho}{L_1})} 
\label{EQ1}
\end{equation}
which contains a constant term $\Sigma {\rm Af}_0$ representing the
non-sublimating (permanent) dust component, and just one decaying term
$\Sigma {\rm Af}_1$ representing sublimating grains and characterized by the
length-scale $L_1$. The fit achieved for 9P is very good and shows
length-scales L$_1$ of similar value for all three near-IR bands,
i.e. $6300 \pm 160$\,km. Using the fixed length-scale 6\,300\,km, the
fitting procedure was done one more time, now for all profiles
including those derived from visible data. The best fit parameters
$\Sigma {\rm Af}_0$ and $\Sigma {\rm Af}_1$, including standard deviations, are
listed in Table \ref{tab_q1}. Again, the fit gives very good results,
with the exceptions of \Uc\ and \Rc\ filters, where $\Sigma {\rm Af}_1$ has
values close to zero.

\begin{table}
\caption{\label{tab_q1} $\Sigma {\rm Af}_0$ and $\Sigma {\rm Af}_1$ best fit results (see text)}
\centering
\begin{tabular}{crr}
\hline 
Band& $\Sigma$Af$_0$(cm)& $\Sigma$Af$_1$(cm)\tabularnewline
\hline
\hline 
\Uc&$345.3 \pm 1.3$&$ 10.7 \pm 4.1$ \\
\Bc&$377.9 \pm 1.3$&$ 62.8 \pm 2.8$ \\
\Rc&$603.5 \pm 0.6$&$  2.5 \pm 2.8$ \\
$J$&$682.0 \pm 0.6$&$251.0 \pm 1.9$ \\
$H$&$758.4 \pm 0.9$&$313.8 \pm 3.1$ \\
\Ks&$792.9 \pm 1.6$&$392.4 \pm 6.3$ \\
\hline
\end{tabular}
\end{table}

\begin{figure}[]
\centering
\vspace*{0.3cm}
\resizebox{!}{5cm}{\includegraphics{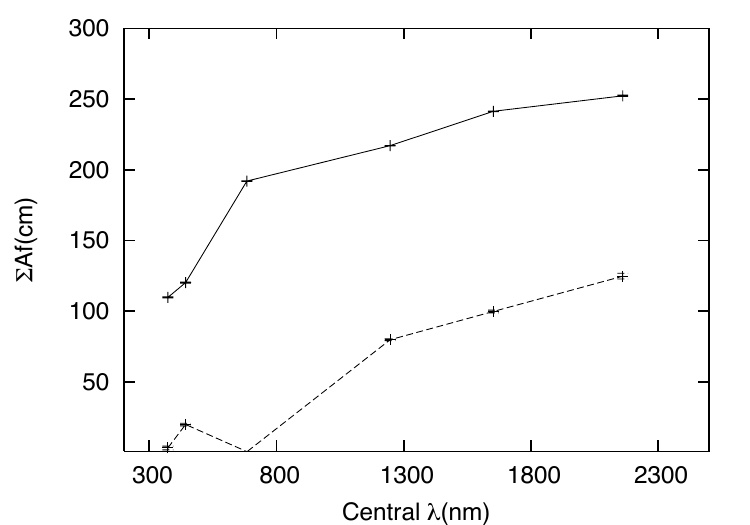}}
\caption{\label{fig_color_quiet}
Wavelength dependence of the permanent (solid line) and
sublimating (dashed line) dust components in the coma of comet 9P
before the impact.  Error bars of the fit parameters $\Sigma$Af$_0$
and $\Sigma$Af$_1$ in eq. \ref{EQ1} are also plotted, but are too
small to be easily seen at the plotted data points.}
\end{figure}

Interesting trends appear when plotting the wavelength dependence of
the decaying ($\Sigma$Af$_1$) and the constant ($\Sigma$Af$_0$) terms
of the fits (see Fig. \ref{fig_color_quiet}). In the near-IR the
constant term varies only by a factor of 1.16 going from $J$ to \Ks,
while the decaying term changes by a factor of 1.55. This finding may
indicate that the two solid components are of a very different nature:
one (the permanent one) composed of refractory grains and the other
(the decaying one) made of sublimating grains (or dust covered by
sublimating material) that scatter very efficiently in the near-IR,
but are inefficient scatterers in the visible light. As for the comet
WM$_1$, taking into account that the length-scale for the density was
about 25\,\% longer than that for the column density, and assuming an
outflow velocity of about 0.2\,\kms, the lifetime of the sublimating
material was of the order of 40000 s (about 11 hours). Assuming that
the sublimation is driven directly by the radiation, the scalelength
and lifetime scale as the square of the heliocentric distance,
r$_h$. Then the density length-scale and the lifetime at r$_h$ = 1\,AU
should be 3\,500\,km and 17\,800\,s ($\approx 5$\,h).  That lifetime
differs from those found for the volatile dust grains in WM$_1$,
which, assuming as well an outflow velocity equal to 0.2\,\kms, were
61\,000\,s ($\approx 17$\,h) and 4700\,s ($\approx$ 1.3\,h) at the
heliocentric distance of that comet (1.2\,AU). They scaled to 42000\,s
($\approx 12$\,h) and 3300\,s (54\,min) at 1\,AU.  The grain
sublimation may not depend directly on the solar irradiation, but may
depend indirectly on it, through the grain temperature. In this case
the lifetime and scalelength scale in a more complicated way with
r$_h$, dependending on grain size and thermal degradation of the
element \citep[see, e.g.,][]{Cottin2004}.

The spectral scattering properties of the sublimating grains in WM$_1$
are different from those found here: both WM$_1$ components have a
good scattering efficiency in $R$ and $I$, and the long lifetime
component has a scattering efficiency similar to the normal dust. The
sublimating component in 9P almost does not scatter in the visible.
This means that, even if the phenomena may be similar for the two
comets, the nature of the grains must be different.

\begin{figure}[]
\centering
\resizebox{!}{6cm}{\includegraphics{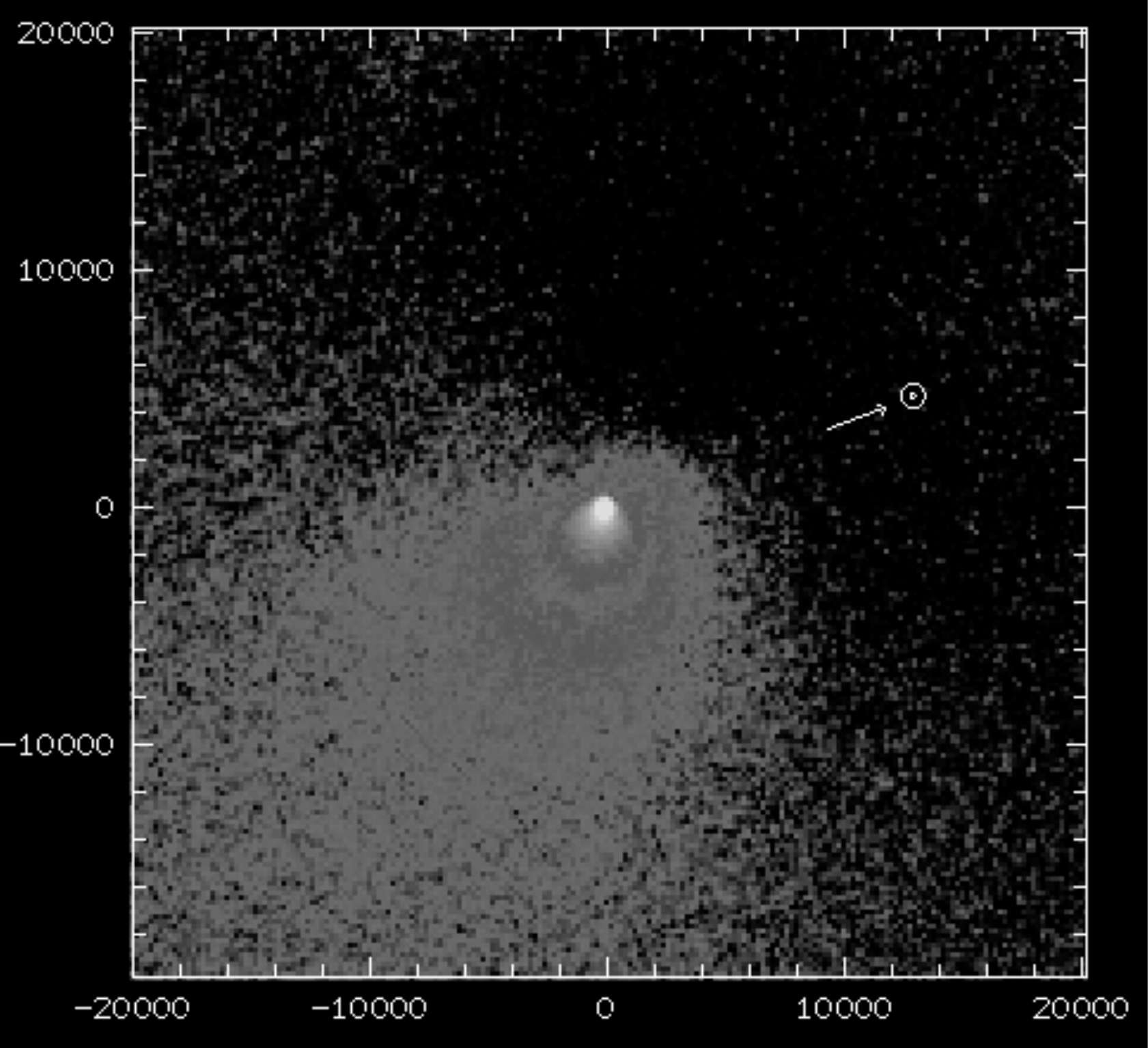}}
\caption{\label{fig_subl_comp} 
Difference image of the "quiet" comet minus a synthetic one,
as computed from a \Ks\ filter exposure of the comet on July 2-3,
2005 following the procedure described in the text. North is up, East
to the left. The Sun direction is indicated. The FoV of the image is $40\,000
\times 40\,000$\,km$^2$ and intensity in Af (0-15 $\times 10^{-7}$),
with a logarithmic look-up-table. The image shows an enhanced flux level
in the southern coma hemisphere due to the sublimating dust and above
average activity from isolated regions on the nucleus (jets\&fans). It
appears as if the sublimating grains are emitted between PA 100 and
200\degr.}
\end{figure}
In order to investigate the origin of the sublimating dust component,
we have tried to determine its 2D distribution in the coma by
subtracting a radially symmetric artificial coma image from each near-IR
filter image. The artificial coma image was computed using the
parameters for the permanent dust component in the respective fit to
the original image. Figure~\ref{fig_subl_comp} shows an example for the
difference image in the \Ks\ filter. The non-uniform flux distribution
in the images suggests that the sublimating dust is more confined in
the coma sector defined approximately by PA = 100\degr ~ to
200\degr. Given the indicated PA range, the appearance of sublimating
dust seems to correlate with the nucleocentric surface regions of
enhanced activity since various jet and fan structures are detectable
in the same coma region (see \citeauthor{Lara2006}
\citeyear{Lara2006} and \citeauthor{Boehnhardt2007} \citeyear{Boehnhardt2007}). However,
the jets and fans are also present in the difference images described
above since the subtraction of a circular symmetric comet image does not cancel any
asymmetries in the coma flux distribution. Thus, it is not possible to
fully disentangle the jet and fan structures from the sublimating dust
component, except that the former have much lower intensity levels
than those of the fading grains.

\subsection{The ejecta cloud}

\subsubsection{Geometry of the ejecta cloud}
\begin{figure*}[]
\centering
\resizebox{!}{6cm}{\includegraphics{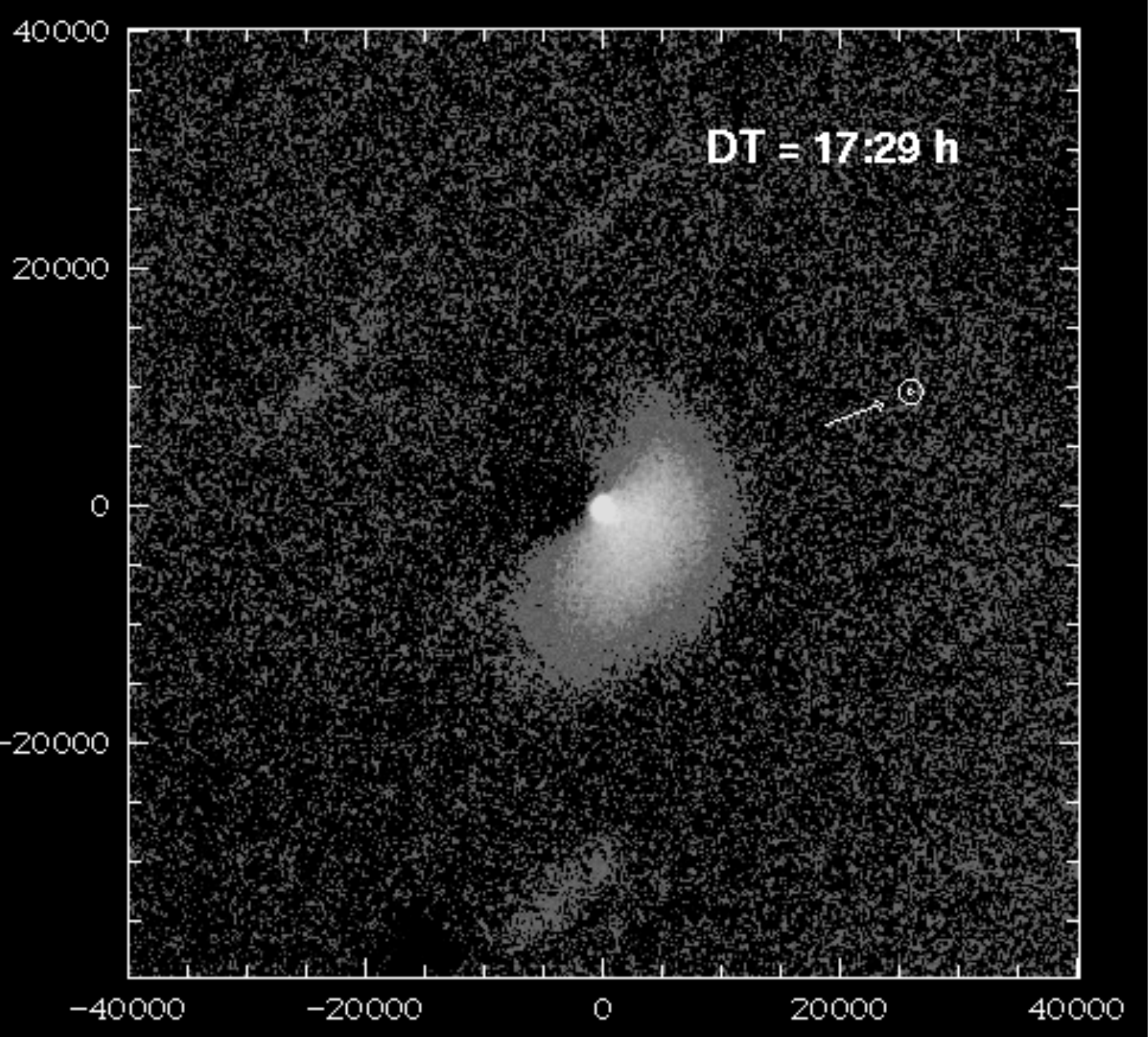}}
\resizebox{!}{6cm}{\includegraphics{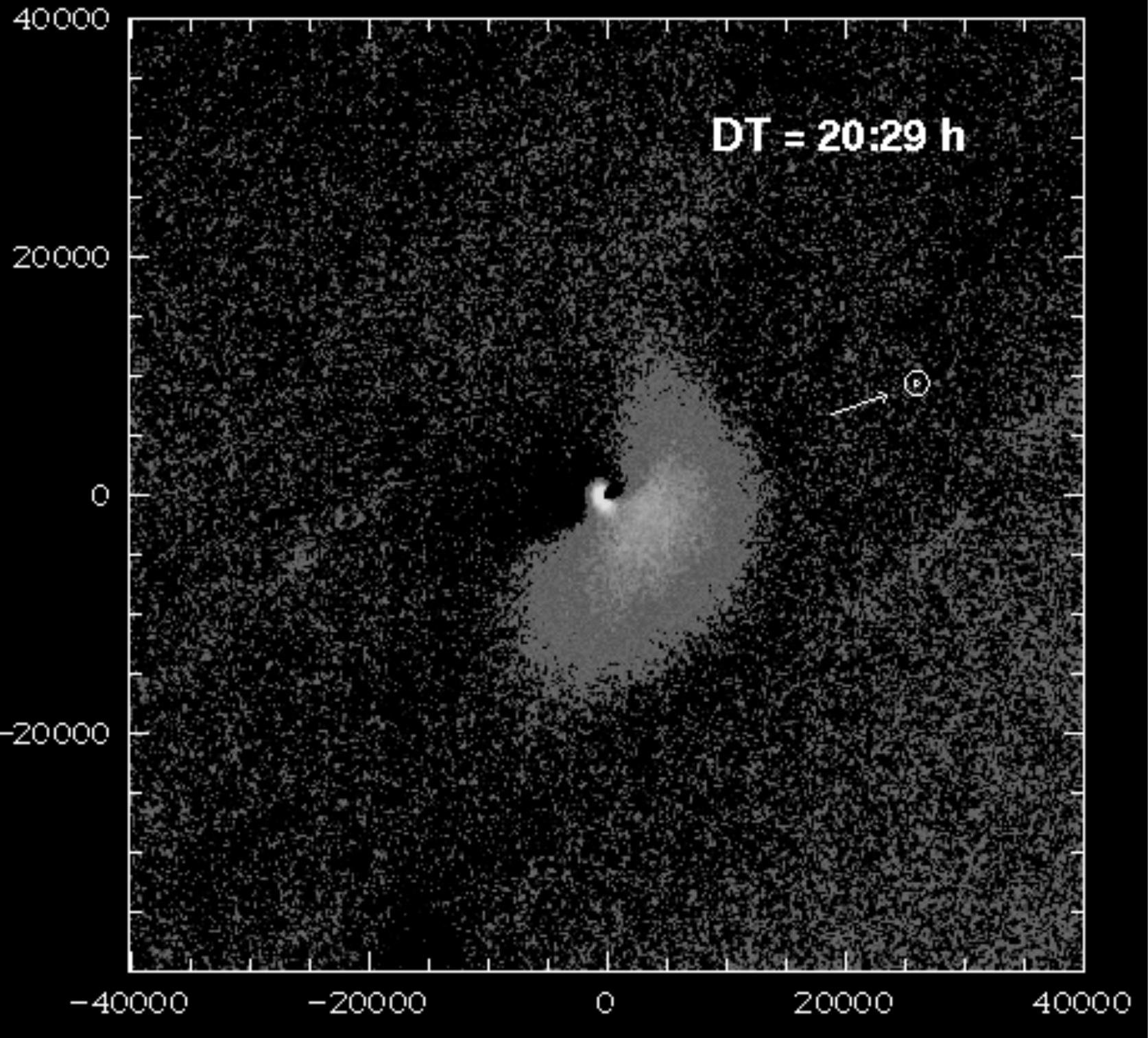}}
\caption{\label{fig_C9P_dif1}
Processed $J$ filter images of the ejecta cloud recorded about
17 and 20 hours after the impact. The "quiet" comet is removed from
the post-impact exposures. The intensity scale is in Af (0--3 $\times
10^{-7}$), with a logarithmic look-up-table.  North is up, East to the
left, the field of view (FOV) at the comet is $80\,000 \times
80\,000$\,km$^2$, the comet nucleus position is in the center of the
sub-panels. The Sun direction is indicated by the arrow. The cloud 20\,h
after the impact is less bright than 3 hours earlier. The "hole" at
the center is an artifact from the subtraction processing of the two
exposures pre- and post-impact which had different seeing. Seeing
variations do not affect the cloud intensity outside of the typical
seeing radius of 1-2 arcsec, i.e. about $700-1\,400$\,km of the
nucleocentric distance.}
\end{figure*}

For the study of the various effects produced by the DI event, the
signal of normal activity was removed from the post-impact images of
9P by simple subtraction of the image of the "quiet" comet, taken
through the same filter. This processing should remove the non-impact
comet coma without introducing new unwanted features from day-to-day
variability, since the normal activity of 9P displayed a rather
steady-state appearance. The expanding cloud of solid particles is
clearly noticeable in the visible and near-IR images until at least
July 6-7, 2005 (three days after DI).  Fig. \ref{fig_C9P_dif1} shows
the ejecta cloud in $J$ band as seen 17:29 and 20:29 hours after the
impact. It can be seen in the figure that the cloud is initially
expanding into the coma sector between PA of 120 to 345\degr. The time
evolution of the cloud expansion can be characterized by using visible
broadband imaging \citep{Boehnhardt2007}.

\subsubsection{Ejecta dust production} 
\begin{figure*}[]
\centering
\vspace*{0.3cm}
\resizebox{!}{5cm}{\includegraphics{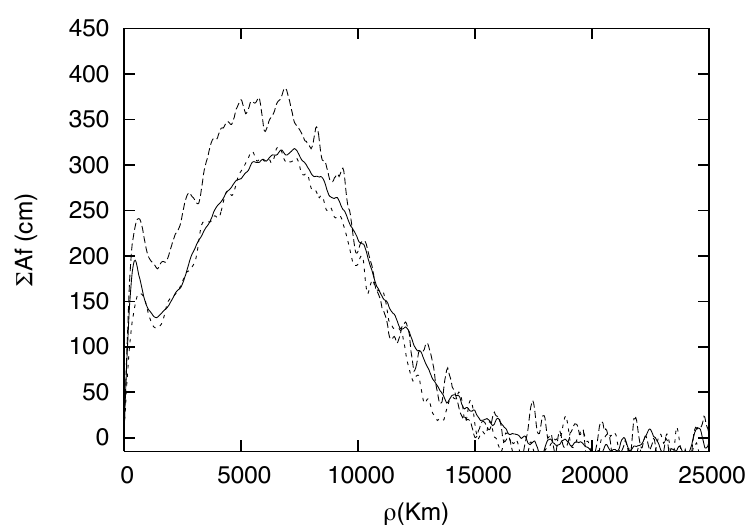}}
\resizebox{!}{5cm}{\includegraphics{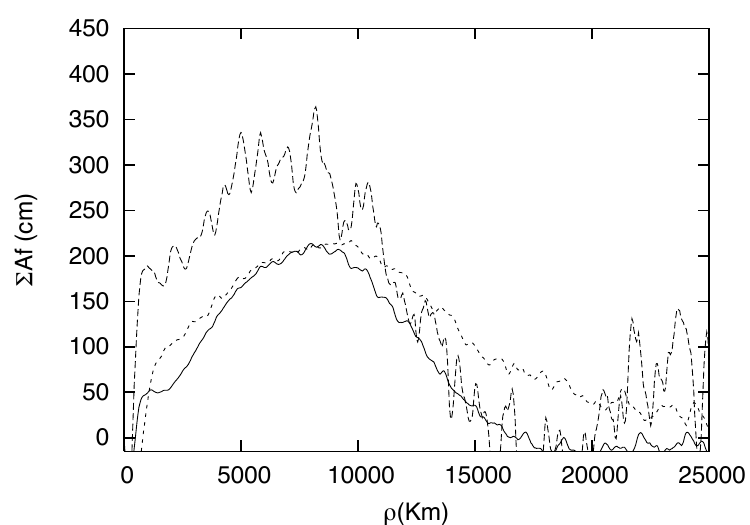}}
\caption{\label{fig_cloud_rho}
$\Sigma$Af profiles of the ejecta cloud vs $\rho$ measured on
night 4-5 July 2005 as a function of $\rho$. Left panel: $J$ (solid
line), $H$ (dotted line) and \Ks\ (long dash line) profiles from
observations taken 17:29 h after the impact. Right panel: $J$ (solid
line), \Ks\ (long dash line) and \Rc\ (dotted line) $\simeq$ 20:29 h
(22:32 h for \Rc) post-impact. Note the changes in the $J$ filter
profiles between 17:29 and 20:29 h post-impact time. }
\end{figure*}

Integrating the Af in the images difference over the position angle
range of the initial ejecta cloud (PA=120--345\degr), we obtain the
$\Sigma$Af profile of the cloud vs $\rho$. The $\Sigma$Af profiles
determined in the \Rc, $J$, $H$ and \Ks\ filters for the night just
after the impact are shown in Fig. \ref{fig_cloud_rho}. Note that the
different extension of the cloud profiles in the figure is due to the
different observing epochs during which the cloud was expanding in the
field of view.

By integrating these profiles over $\rho$, we can obtain the total
scattering cross section (SA) of the dust ejecta, SA = $\int{({\Sigma
{\rm Af}})d \rho}$, i.e., the albedo at the phase angle of the comet
multiplied by the total geometric grain cross section. SA provides
useful information to evaluate the number and the intrinsic color of
the particles produced by the impact and their evolution with
time. For the first observations of the first night (July 4-5, 2005)
after the impact, we measured the SA to be 27.6, 27.3 and 34.6\,km$^2$
in $J$, $H$ and \Ks\ respectively. Assuming that the scattering
properties of the ejecta grains are the same as those of the
refractory component in the pre-impact coma, it is possible to
estimate the time interval TI necessary to produce the same amount of
dust by the normal activity: TI = $\frac{SA}{v \Sigma {\rm Af}_0}$,
where v is the mean outflow velocity, and $\Sigma$Af$_0$ are the
values obtained for the normal activity. To estimate the order of
magnitude, any possible differences in the scattering properties of
the dust grains were ignored and a $v = 0.2$\,\kms\ was assumed. Then
TI is about $5-6$ hours (depending on the filter). For the lower
velocity $v = 0.1$\,\kms, the equivalent duration of normal dust
production is doubled.  We conclude that the amount of dust produced
by DI and detectable in the near-IR about 17-20 h post-impact time is
equal to the amount of the dust produced during a few hours (maximum
half a day) of normal activity of the comet just before the impact.

Results for the effective scattering cross section SA in km$^2$,
estimated from the dust filter images available to us, are tabulated
in Table \ref{tabintsaf}. The relative error for different filters
depends mainly on the respective flux calibration uncertainty that is
estimated to be of the order of 10\,\%. The relative error for
different measurements with the same filter does not depend on this
calibration, because during the data reduction the $\Sigma$Af profiles
of the ejecta cloud images were checked to match the reference profile
of the "quiet" comet at nucleus distances beyond 30\,000\,km. Thus,
the relative uncertainty in $\Sigma$Af for different post-impact
epochs depends only on the accuracy of the matching of the "quiet"
comet profiles. As seen in Figs. \ref{fig_C9P_dif1} and
\ref{fig_cloud_rho}, a rather accurate match is achieved and, hence,
the relative error of $\Sigma$Af is evaluated to be less that
5\,\%. The values shown in the table are in agreement with the $33 \pm
3$\,km$^2$ obtained in the visible range from Rosetta/Osiris
observations  about 40 minutes after the
impact \citep{Kueppers2005}. 

\begin{table}
\centering
\caption{\label{tabintsaf}
Effective cross section of the ejecta grains in km$^2$ vs. time $t-t_0$ 
after impact for different dust filters. 'Effective cross section' is defined 
in the text.}
\centering
\begin{tabular}{ccccc}
\hline
$t-t_0$ (hh:mm) & SA(\Rc) & SA($J$)  & SA($H$) & SA(\Ks) \\ 
\hline
\hline
17:29 &     & 27.6 & 27.3 & 34.6 \\ 
20:29 &     & 17.9 &     & 33.0 \\ 
22:32 & 27.3 &     &     &      \\
\hline
\end{tabular}
\end{table}

\subsubsection{Ejecta velocities} 
Contrary to the natural outbursts in comets, in the case of DI event
the exact starting time of the formation of the ejecta cloud is very
well known. Hence, assuming that the major dust production by DI was
short-term (as suggested by the fly-by spacecraft imaging; see
\citeauthor{AHearn2005} \citeyear{AHearn2005}), the radial profiles of
the ejecta cloud would also reflect the expansion velocity
distribution of the cloud particles. The solar radiation pressure
applies some acceleration $\frac{dv}{dt}$ to the particles, which is
inversely proportional to the particle radius, $a$.  It has been shown
that some time between 1 and 2 days after impact the dust grains, as
observed in the visible, reached the turning point in their motion in
the sunward direction \citep{Boehnhardt2007} due to the solar radiation
pressure.

\begin{figure}[]
\centering
\resizebox{!}{5cm}{\includegraphics{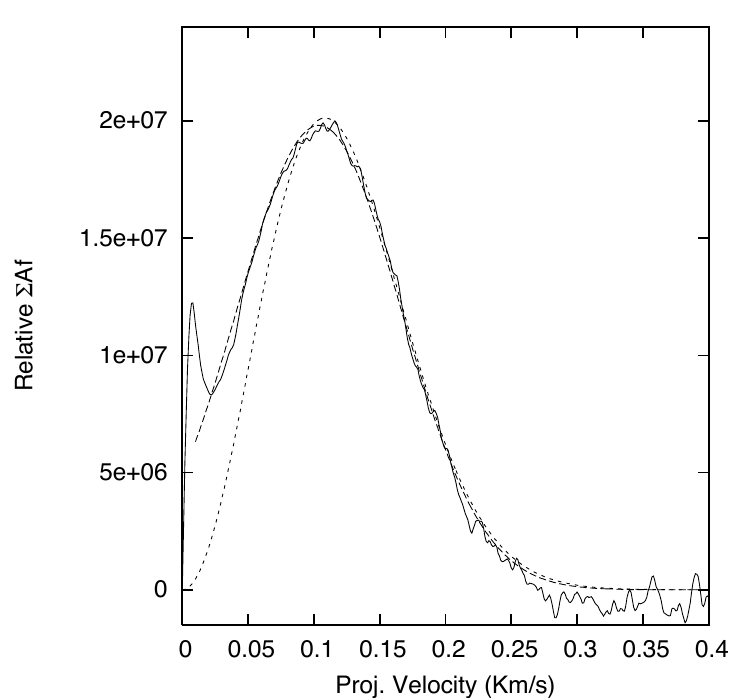}}
\caption{\label{fig_cloud_v}
The $\Sigma$Af(v) profile, which is proportional to the mean
  projected velocity distribution of the ejecta dust, as obtained from
  $J$ filter imaging during night 4-5 July 2005. The measured profile
  (solid line) is compared to Gaussian (dashed line) and Maxwellian
  distributions (dotted lines). }
\end{figure}

So, depending on the grain size, the radial profiles
$\Sigma$Af($\rho$) of the ejecta cloud show a memory of the initial
projected velocities of the dust after the ejection from the nucleus and
possible further acceleration in the near-nucleus zone by the ejecta
gas and/or normal gas release activity.  Thus, $\Sigma$Af($\rho$)
divided by the elapse time since the impact results in some
distribution function $\Sigma$Af(v) where v is the mean 
velocity of the dust grain at the respective projected nucleocentric
distance in the cloud.  A typical mean velocity distribution is shown
in Fig. \ref{fig_cloud_v}. As pointed out by \citet{Kueppers2005} and
\citet{Jorda2007} the radial profile of the eject cloud can be well
fitted with a Gaussian function. In Fig. \ref{fig_cloud_v}, the
Gaussian function that provides the best fit to the velocity profile
is overplotted. The agreement is very good for all the filters. For
comparison, a Maxwellian function is also shown. Here the agreement
with the data is good in the leading part of the function, but is poor for
small velocities.

Because of the radiation pressure (see above), the velocity
distribution is not the same in all directions; the mean velocity is
lower in the Sun direction than in other directions.  To
study this effect, the cloud images have been divided in three
sectors: S1, S2, and S3 defined by their position angle ranges
145--204\degr, 205--264\degr and 265--325\degr, respectively. The
velocity profiles were then fitted with a Gaussian function.  Results
are shown in Table \ref{tabvel}, where the mean velocity $\bar{V}$ and
its FWHM is given for nights July 4-5 and 5-6.  The table does not
indicate the errors of the fitting since they are very small, of the
order of $\pm 2$\,\kms\ and $\pm 4$\,\kms\ for the first and second
night after the impact, respectively.

\begin{table}
\centering
\caption{\label{tabvel} 
Average projected velocities $\bar{V}$ and FWHM, in m$s^{-1}$, as
determined in the three sectors S1, S2 and S3 of the dust ejecta cloud
(see text) for different filters.}
\centering
\begin{tabular}{cccccccc}
\hline
     &$t-t_0$& \multicolumn{2}{ c|}{Sector S1}& \multicolumn{2}{c|}{Sector S2}&\multicolumn{2}{c}{Sector S3}\\ 
Band & (hh:mm) & $\bar{V}$  & FWHM & $\bar{V}$  & FWHM  & $\bar{V}$ & FWHM \\ 
\hline
\hline
 \Rc   & 23:32     & 138 & 95   & 103 & 61    & 87 & 50   \\ 
 $J$   & 17:29     & 120 & 69   & 100 & 58    & 98 & 51   \\ 
 $J$   & 20:29     & 123 & 57   & 101 & 48    &100 & 40   \\ 
 $J$   & 45.32     & 134 & 48   &  98 & 44    & 93 & 42   \\ 
 $H$   & 17:16     & 117 & 69   &  98 & 56    & 98 & 49   \\ 
 $H$   & 45:73     & 133 & 58   & 102 & 53    & 76 & 41   \\ 
\Ks    & 17:03     & 113 & 71   &  95 & 63    & 91 & 50   \\ 
\Ks    & 20:13     & 143 & 87   &  84 & 52    & 76 & 41   \\ 
\Ks    & 45:55     & 122 & 62   & 115 & 54    & 97 & 41   \\ 
\hline
\end{tabular}
\end{table}

Table \ref{tabvel} indicates the following: (1) $\bar{V}$, as measured in the near-IR, is
independent of the filter, (2)
$\bar{V}$ depends on the sector, it is slowest in S3, but shows very
similar values for the three near-IR filters, (3)
$\bar{V}$ in sector S3 slightly decreases with time, while in sector
S1 it increases with time.

This picture can be explained by the solar radiation pressure, since
the projected Sun direction at PA $\simeq$ 290\degr\  falls almost
exactly in the middle of sector S3.  The average values of the mean
velocities, as in the near-IR, are the following: $\bar{V}_{S1}$ =
123$\pm$12, $\bar{V}_{S2}$ = 96$\pm$7, $\bar{V}_{S3}$ =
93$\pm$10\,\ms\ for the night July 4-5 and $\bar{V}_{S1}$ = 130$\pm$7,
$\bar{V}_{S2}$ = 105$\pm$9, $\bar{V}_{S3} = 89 \pm 11$\,\ms\ for the
subsequent night.  Using simple physical considerations, the projected
distance covered by a grain in the S3 sector ($\simeq$Sun direction)
is $S_{S3}(t) = V_{ej} t-\frac{1}{2}\frac{dv}{dt} \times t^2$ and, in
the sector S1, ($\simeq$60\degr\  from the antisun direction) $S_{S1}(t) =
V_{ej} t + \frac{1}{2}\frac{dv}{dt} cos(60) \times t^2$. With simple
algebraic operations the V$_{ej}$ results are $(\bar{V}_{S3}(t)+2
\bar{V}_{S1}(t))/3$. Using the average values given above, the results
are $V_{ej} = 113\pm 16$ and $116\pm 16$\,\ms, for the nights July 4-5
and 5-6, respectively. The typical FWHM is of the order of 75 \ms~
for both nights. In the visible the eject velocity is estimated to be 
of the order of 120 \ms. Since only a single observation of this kind is
available to us, we adopt an uncertainty similar to those in the near-IR.

\subsubsection{Effective scattering cross section}
A surprising finding is the rapid decrease in the effective $J$
scattering cross section SA of the ejecta cloud by 35\,\% during the
time interval from 17:29 to 20:29 hours after the impact as seen in
Table \ref{tabintsaf}.  This is also noticeable in
Fig. \ref{fig_C9P_dif1}, which demonstrates that the cloud at
20:29-hours post-impact is significantly fainter than three hours
earlier (17:29 h post-impact) even though it has almost the same
extension. Also, the change in the radial profiles $\Sigma$Af of the
$J$ filter data 17:29 and 20:29 h after the impact shown in Fig.
\ref{fig_cloud_rho} confirms the change in the scattering of the
ejecta particles. This cannot be due to calibration problems, because
SA is computed using the difference between the impact images minus
the "quiet" comet: calibration problems would give values different
from zero in regions far from the cloud. However, this is not the
case, as can be seen in Figs.  \ref{fig_C9P_dif1} and
\ref{fig_cloud_rho}.

The cloud color 17:29h after the impact is almost 'gray' from $J$ to
$H$, but its scattering efficiency increases by 25\,\% in \Ks. Three
hours later (20:29 h post-impact), the scattering efficiency increases
by 84\,\% between $J$ and \Ks\ (observations in $H$ were not taken
because of the lack of time). Thus, the reduction in the effective
scattering cross section of the dust cloud in the near-IR between
17:29 and 20:29 h after the DI event was accompanied by a strong
reddening of the dust.

\begin{figure*}[]
\centering
\resizebox{!}{4.5cm}{\includegraphics{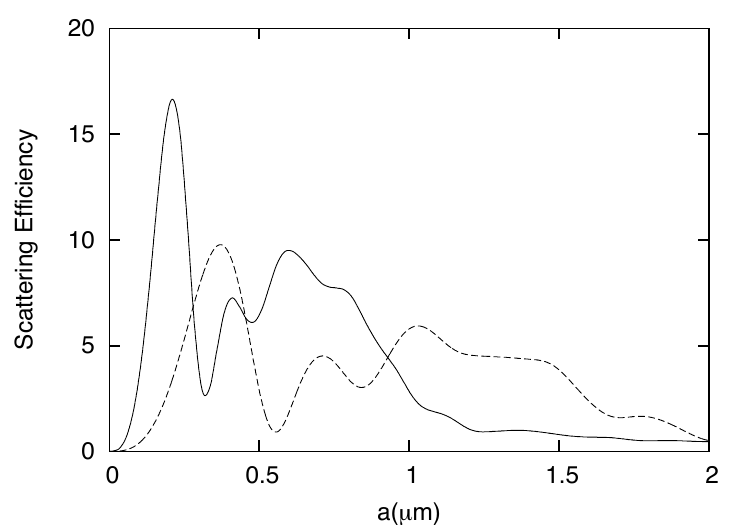}}
\resizebox{!}{4.5cm}{\includegraphics{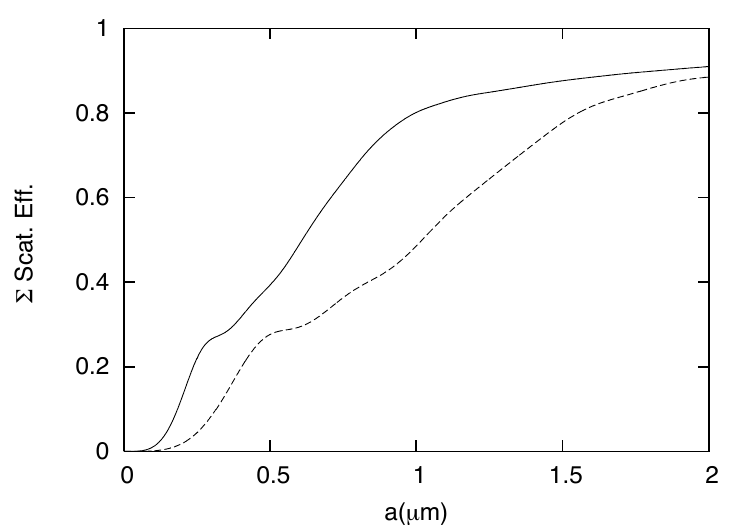}}
\caption{\label{fig_scattering}
Differential (left panel) and relative cumulative (right panel)
scattering efficiency of a cloud of particles, as a function of their radius
(see text), as observed in the filter $J$ (solid line) and in \Ks\ (dotted
line). }
\end{figure*}

To study this, the scattering efficiency for the three near-IR filters
as a function of particle radius, $a$, was modeled. In the model,
spherical particles were assumed with a power-law size distribution of
power --3.1 as derived by \citet{Jorda2007} using data obtained 40 min
after the impact.  The refractive index was set to 1.65+i0.062, that
represents a mixture of silicates and organics typical for comets
(\citet{Jessberger1988}.  Results are shown in
Fig. \ref{fig_scattering}: the left panel gives the scattering
efficiency for $J$ and \Ks\ bands as a function of the particle
radius, while the right panel gives the cumulative scattering
efficiency of the particles, normalized to 1 at large sizes (5
$\mu$m). From the figures it is possible to see that the particles
with radius less that 0.1 $\mu$m give a negligible contribution to the
scattering and that 80\,\% of the total scattering is reached for the
particles with $a \leq$ 1 and $\leq$ 1.5 $\mu$m for $J$ and \Ks,
respectively. Note that the particles with $a \leq$ 0.92 $\mu$m are
mainly responsible for the greatest difference between the cumulative
scattering in $J$ and \Ks. Indeed, these particles provide 77\,\% of
the total scattering in $J$ and only 44\,\% in \Ks. From these results
it is possible to derive an important conclusion: if the destruction
or sublimation of particles with $a \leq$ 0.92 $\mu$m is responsible
for the above-mentioned decrease of SA in the $J$ band, a destruction
of 50\,\% of such particles should decrease the SA by 35\,\%. However,
since these particles ($a \leq 0.92 \mu$m) contribute also to the
scattering in \Ks\ (44\,\%), this destruction should also produce a
22\,\% decrease of the SA in \Ks filter. Instead, the measured
decrease for this filter is only 5\,\%.

A possible conclusion is that the observed particles do not obey a
power law size distribution. The results obtained by the DI
spacecraft spectrometer indicates that the original DI size
distribution was dominated by particles of a few microns in size
(\citeauthor{AHearn2005} \citeyear{AHearn2005} and supporting
on-line material of \citeauthor{Lisse2006} \citeyear{Lisse2006}). The
difference between \citet{Jorda2007} and this result may be due to a
poor sensitivity of the Jorda et al. measurements to particles of
size larger than 1 $\mu$m as is expected for the measurements
in the visual, as pointed out in Jorda et al.. Thus, the particles
of size $a \approx$ 1 $\mu$m could dominate in the ejecta cloud 17:29
h after the impact and produce a significant contribution to the
radiation measured in $J$ filter.  It is well known that small grains
sublimate faster than large ones because they warm up more efficiently
than the larger ones, since the absorption of the solar radiation is
proportional to their area and the heating is inverse proportional to
their volume. This results in the 1/$a$ law \citep[see,
e.g.][]{Lamy1974}), i.e. the smaller the particles are the faster they
sublimate. This means that 1 $\mu$m particles may sublimate faster and be
eliminated from the size distribution more efficiently than larger
particles. Thus, 20:29 h after the impact the maximum of
the size distribution shifts to larger particles. This leaves the $J$
band without the most efficient contributors whereas the situation
the \Ks\ band remains almost unchanged. This scenario, consistent
with the in situ data, gives a hope that careful simulations of the
sublimation of Deep Impact ejecta particles may provide some
information about the sublimation rate of the ejecta volatiles, and,
thus, may help to identify it.

An alternative scenario would be a change in the particle composition
which significantly modified the value of the refractive index.
However, this hypothesis cannot explain the observations. First, the
most dramatic changes would be expected shortly after the impact when
the most volatile components of the dust, e.g. ice, sublimate. For the
less volatile components the change should be slow and it is hard to
imagine how such a dramatic change happened between 17:29 and 20:29
hours after the impact. Second, we are not aware of any material
that may be expected in comets that has a so significant difference in
its optical properties between $J$ and \Ks\ filters.

\subsubsection{Color gradient}

Due to the different orientation of the grain velocity vector to the
radiation pressure force, radiation pressure has different effects on
the dust grains in sectors S1, S2, and S3 of the ejecta cloud. This
results in different sorting of dust particles by size for different
sectors which may appear as a difference in the dust colors. From the
$\Sigma$Af($\rho$) profiles the color distribution of the cloud is
computed and the results are averaged over the PA range of the three
considered sectors. Since only the near-IR images had sufficient
signal-to-noise ratio within the nucleocentric distance range of
2\,000--9\,000\,km, and only $J$ and \Ks\ observations were performed
during the first night after impact, only the $J-\Ks$ color (C) is
computed. The results, in percent per nm and per 1\,000\,km, for the
night July 4-5, 2005 show that the ejecta cloud becomes 'bluer' with
larger distance from the nucleus. From the images $\simeq$ 17:29 h
after the impact we have found the following:

$C_{S1} = (4.1\pm0.1) - (0.31\pm0.02) \% nm^{-1} (1000 {\rm km})^{-1}$

$C_{S2} = (3.8\pm0.1) - (0.36\pm0.02) \% nm^{-1} (1000 {\rm km})^{-1}$

$C_{S3} = (5.0\pm0.1) - (0.51\pm0.02) \% nm^{-1} (1000 {\rm km})^{-1}$

Thus, the grains closer to the nucleus scattered about 40\,\% (4\,\%
$\times$ the difference of the central wavelength of \Ks\ and $J$)
more efficiently in \Ks\ than in $J$. However, they scatter with
almost the same efficiency in sector S3, at about 10\,000\,km from the
nucleus. The strong difference in the gradients between the three
sectors can be explained by the effect of the solar radiation pressure
mentioned above.

Three hours later (i.e. 20:29 h after the impact) we find:

$C_{S1} = (~7.5\pm0.3) - (0.28\pm0.06) \% nm^{-1} (1000 {\rm km})^{-1}$

$C_{S2} = (10.7\pm0.3) - (1.06\pm0.02) \% nm^{-1} (1000 {\rm km})^{-1}$

$C_{S3} = (15.5\pm0.3) - (1.76\pm0.02) \% nm^{-1} (1000 {\rm km})^{-1}$
 
\begin{figure}[]
\centering
\resizebox{!}{5cm}{\includegraphics{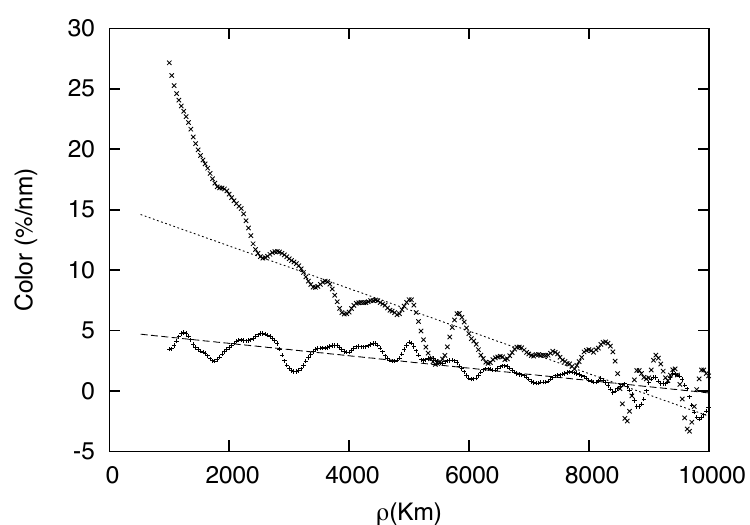}}
\caption{\label{fig_color_rho}
$\Ks - J$ color reddening in \% nm$^{-1}$ of the ejecta cloud in
sector S3 as a function of the projected distance from the
nucleus. Lower graph is obtained from observations taken 17:29 hours
after the impact, the upper graph applies to 20:29 hours after
DI. Dotted and dashed lines show linear fits to the
measurements. While the color remained almost the same at distances
$\rho = 8\,000-10\,000$\,km in the cloud, at smaller distances both
the color and its spatial gradient increased significantly during the 3 hours
between the two observation sequences.  Note also the non-linear
increase of the reddening function for projected distances below
3\,000\,km, as seen in the data obtained 20:29 hours after impact.}
\end{figure}

As is expected from decrease in the effective
scattering cross section in the $J$ filter, a significant increase in the
dust reddening took place close to the nucleus. The scattering
efficiency of the grains close to the nucleus became a factor 2-3
higher in \Ks\ than in $J$. We also notice a strong change in the
spatial gradients of the dust reddening: while the gradient in S1
changes only by about 25\,\%, in the two other sectors the change is
significantly more evident, e.g. by factor of 3 for S3. Moreover, the
increase is non-linear in the inner part of the cloud, as can be
seen in Fig. \ref{fig_color_rho}.

These facts cannot be explained by the solar radiation pressure only.
Also, from our estimation of the speed of particles, the three hours
between our observations cannot bring a significant number of new
particles: the particles can travel only $\approx 20$\,km farther from
the nucleus. Thus, 17:29 hours after the impact and 3 hours later we
observed particles of the same origin. However, their properties
and/or size distribution had changed dramatically. This is confirmed
by the above change in the scattering cross sections. The changes in
color likely also resulted from the sublimation of particles which had
some specific, non-power-law size distribution.  This dominance of
particles of different sizes is what determines the color and its
gradient. The difference in the color gradients may indicate the
efficiency of sublimation whose rate increases as particle size
decreases (see Section 4.2.4).

\section{Discussion}
An important result of these observations has been the discovery of the
sublimating component in the coma of the so-called "quiet" comet. This
kind of fading grain has been found in comet C/2000 WM$_1$, where
two components, one with a short and the other with a long lifetime, were
found when the comet was at 1.2\,AU from the Sun. These components
were interpreted as organic grains or refractory grains embedded in
organic matter that sublimated while heated by the solar radiation.
The sublimating component discovered in comet 9P seems to be different
from both sublimating components found in comet C/2000 WM$_1$. It
scatters very efficiently in the near-IR, but it does not scatter at
all in the visible.  Scaled to 1\,AU, its length-scale is of the order
of 3\,500\,km and, assuming an outflow velocity of 0.2\,\kms, the
lifetime is 18\,000\,s (5h). Note that \citet{Cottin2004} suggested
that refractory organic grains, namely polyoxymethylene (POM), may be responsible for the distributed source of formaldehyde, observed
in several comets. They computed the scalength and lifetime of the POM
grains assuming photolysis by solar radiation and thermal
sublimation. With a temperature of
grains of 350\,K, they computed a POM length-scale of the same order of
magnitude as that measured in 9P, depending very little on grain
size. It varied from 3\,300 to 7\,100\,km for sizes going from 0.1 to
10\,$\mu$m.

The following
results are obtained for the ejecta cloud produced by the impact :
\begin{itemize}
\item {The total amount of the dust, multiplied by the albedo, covers
 a surface of about 30\,km$^{2}$ about 17\,h after the impact, but it
 drops dramatically for the $J$ band 3\,h later; }
\item{ The velocity distribution of the solid components
 had a Gaussian distribution with an average ejected velocity equal to 115$\pm$16 \ms with a FWHM of the order of 75\ms;}
\item{ The velocities in the projected
direction of the Sun are smaller than those in other directions and have
similar values for the near-IR  bands. Those 
in the visible have larger values than the near-IR ones;}
\item {From the observations 17:29 h hours after the impact, the
near-IR color  of the grains close to the nucleus was found to be very red and shows
a strong gradient with the nucleocentric distance with the highest
values in the Sun direction. Three hours later, the near-IR color
becomes even more red and the gradient with $\rho$ in the Sun
direction increases by a factor $\simeq 3$.}
\end{itemize}

The Gaussian velocity distribution of the particles is puzzling. In
the case of gas drag produced by an explosive event that has a
timescale shorter than the acceleration time scale, the velocity of a
particle with radius $a$ should change as $a^{-1}$. This, combined
with a power-law size distribution of power equal to --3.1
\citep{Jorda2007} would produce a velocity distribution very different
from a Gaussian one. If the lifetime of the explosive event was longer
than the timescale of the acceleration the ejection velocity
distribution with particle sizes would follow a more complex law, but
always size dependent \citep[see, e.g.][]{Gombosi1986}).  In the case
of a natural outburst the velocity distribution has been found to
follow a Maxwellian one \citep[see, e.g.,][]{Schulz2000}. On the other
hand, a thermal acceleration of the dust, that would give a Maxwellian
distribution, is excluded because it would require a temperature
excessively high to accelerate a grain with the mass of the order of
10$^{-16}$ g at a velocity of a hundred \ms. Models of ejecta with
grains with a power law size distribution, accelerated by gas drag,
give a velocity distribution far from a Gaussian one.  Only in the
case of grains with almost the same size would the velocity
distribution follow that of the grains and it can also became a
Gaussian one.  The average projected velocity found here is in good
agreement with the value of 115 \ms\ obtained by by
\citet{Feldman2007} from HST observations in the visible just after
the impact. It is smaller than those noted by \citet{Meech2005}
(200$\pm$20 \ms) and \cite{Schleicher2005} (220 \ms) who refer to the
velocity of the leading part of the cloud. For the latter we find
250-300 \ms, in fairly good agreement with their values.

However, our result for the Gaussian velocity distribution of the
expanding ejecta cloud differs from that of \citet{Jorda2007} (190\ms\
with FWHM = 150\ms), derived from Osiris observations on-board
Rosetta. This difference cannot be explained with the different
viewing geometry of Rosetta spacecraft with respect to an observer on
Earth (the difference in aspect angle is just 20\degr). The Osiris and
our ESO measurements, obtained at similar wavelengths, are in a good
agreement for the total light scattering area of the dust ejecta (33
km$^2$ by Osiris and 27.3 km$^2$ for our \Rc\ observations). An
explanation for the different velocity of small and faster grains
resulting from dust sublimation a few hours after the impact is thus
very unlikely.

The dramatic changes in the total amount of dust observed in $J$ band
is another very puzzling result.  It is associated with a strong
change in the color of the cloud and its gradient with $\rho$.  A
possible scenario would be a sublimation of the dust grains containing
slow-sublimating volatiles such as organics. Water ice is excluded
because its sublimation timescale is much shorter than 17 \,h
\citep{Hanner1981}.  If the particles of a few micron size dominated in
the original size distribution (as was found from the {\it in-situ}
data) then sublimation of such particles would result in elimination
of the most abundant 1 micron size particles, which are the most
efficient in the $J$ band. This will manifest in a significant
decrease of the brightness in $J$ band as well as by a change in the
dust color. 

This means that the scattering area SA of the grains just after the
impact must have been much larger than that measured at 17:29 h
post-impact, i.e., the quantity of the solid component released by the
impact may have been of an order of magnitude larger than the 5-10h of
normal activity derived in Section 4.2 without considering sublimation
effects. However this is not confirmed by the results of
\citet{Kueppers2005} who found a light scattering cross section of the
cloud in agreement with that found here (see above).  It is not
possible to obtain more information from the results presented here,
but about 17 hours after the impact the size distribution of the
grains seems not to follow a power law (if it ever did), but seems
"monochromatic".

There is no obvious correlation between the
organic grains found in the `quiet' comet and those supposed to be
present in the ejecta cloud.

\section{Conclusions}
From observations of gas emission-free regions of the comet 9P/Tempel\,1
made before and after the Deep Impact event, the scattering
characteristics and the velocity of the ejecta cloud produced by the
impact have been measured. 

Seventeen and a half hours after the impact, the total area covered by
the grains of the ejecta cloud, multiplied by their albedo, was
$27-35$\,km$^2$ in $JH\Ks$ . Three hours later, it dropped to $\approx
18$\,km$^2$ in $J$, but remained almost constant in \Ks. During this
interval of time, the $J-\Ks$ color gradient with the nucleocentric
distance $\rho$ also changed significantly, increasing by a factor of
three in the direction of the Sun.

The projected average velocity of the ejected cloud measured in the
near-IR was $115 \pm 16$\,\ms, and was found to be independent of the
filter used for the observations and the position angle. Its
distribution was very similar to a Gaussian.

It has been shown that all these results cannot be explained assuming
that grain size has a power distribution nor it is possible to assume
that grains are ejected by gas drag.

While the mechanism of grain ejection is difficult to explain, the
behavior afterward can be justified only with the presence of $\approx
1$\,$\mu$m size organic grains that are sublimated by the solar
radiation.

From the pre-impact and late post-impact observations, the presence
of a sublimating component has been detected and interpreted in terms
of organic grains that sublimate because of the solar radiation.

\end{document}